\documentstyle{article}
\begin{document}

\title{Feynman geometry}

\author{Sen Hu and Andrey Losev}

\maketitle

\begin{center}
{\small
School of Mathematical Sciences, University of Science and Technology of China\\
Department of Mathematics, Moscow Higher School of Economics}
\end{center}

\let\thefootnote\relax\footnotetext{Emails: shu@ustc.edu.cn, aslosev2@gmail.com}

\begin{abstract}
  In this paper we introduce a notion of Feynman geometry on which quantum field theories could be properly defined.
  A strong Feynman geometry is a geometry when the vector space of $A_\infty$ structures is finite dimensional. A weak Feynman geometry
  is a geometry when the vector space of $A_\infty$ structures is infinite dimensional while the relevant operators are of trace-class.
  We construct families of Feynman geometries with "continuum" as their limit.
\end{abstract}

\section{Introduction}

Non-Abelian gauge symmetry, as discovered by Yang-Mills, is a basic concept in modern physics. There is a basic principle established by A. Einstein and C. N. Yang: symmetry dictates interactions, e.g. Lagrangians are determined by symmetries. Gauge invariance, conformal invariance and renormalizability of the theory determine the Lagrangian for the standard model. Quantum Yang-Mills theory is now the foundation of most of elementary particle theory, and its predictions have been tested at many experimental laboratories, but its mathematical foundation is still unclear.

Symmetries used by A. Einstein and C. N. Yang follow from differential geometry. In case of A. Einstein theory of general relativity it is a diffeomorphism symmetry that is a fundamental equivalence between smooth manifolds. In case of C. N. Yang theory the gauge symmetry comes from the geometry of principal fiber bundles, and may be understood as diffeomorphisms that preserve bundle structure.

Would these symmetries persist in the quantum world or should they be modified? The answer to this question depends on a more fundamental question: should we consider the geometry of the space-time in QFT the same classical differential geometry or it has to be modified? The following paper discusses this issue and concludes that most probably the geometry has to be modified to what we call Feynman geometry.

There are mainly two approaches in the construction of quantum field theories: Dirac-Segal's approach and Feynman's approach. In Dirac's approach one constructs quantum theories by abandoning classical physical concepts while keeping classical concept of space-time. This approach was pursued by many prominent physicists and mathematicians, namely by Atiyah-Segal-Witten and by Belavin-Polyakov-Zamalochikov, and many fruitful mathematical results were obtained by this approach.

In Feynman's approach it saves many classical concepts such as space-time, trajectories, actions etc. It replaces the minimal action principle in classical mechanics by path integral to calculate various correlation functions. This approach is more familiar to physicists.

Here we propose a new approach to construct quantum gauge field theories by constructing quantum many-body systems over some spaces we call it de Rham type of DGA. As long as we have de Rham type of DGA we could construct all kinds of actions. We demonstrate this in section 4.

The space of de Rham type of DGA are characterized by Feynman geometries. A strong Feynman geometry is a geometry when the vector space of $A_\infty$ structures is finite dimensional. A weak Feynman geometry is a geometry when the vector space of $A_\infty$ structures is infinite dimensional while the relevant operators are of trace-class. This notion unifies several important works, including lattice theory, fuzzy supersphere, Feynman geometry of Hodge type (momentum cutoff), Mnev's construction of $A_\infty$ structure via Whitney forms, Costello's construction of $A_\infty$ structures via homotopy associativity, and Zwiebach's string field theory. We construct families of Feynman geometries with "continuum" as their limit.

We propose to construct quantum field theories over an continuum as limit of a family of theories of Feynman geometries. The advantage of such a construction is its non-perturbative feature.

\section{Dirac-Segal and Feynman approaches to QFT}

The main question of theoretical physics is what is the definition of Quantum Field Theory. Despite of many attempts this question still does not have a completely satisfactory solution. These attempts may be separated into two different
approaches that we may call Dirac-Segal approach and Feynman approach.

\subsection{Dirac-Segal approach and its classical limit}

In the Dirac-Segal approach the space-time is considered a very classical Riemann geometry - smooth manifolds with the boundary, but "physics" has nothing to do with classical physics - instead of fields and integrals over the space of fields we have functors from the category of cobordisms (equipped with some geometrical structure) into category of vector spaces.
We put Dirac name here on the first place since he was first to understand that strongly quantum mechanics where
cobordisms are just intervals, geometrical data - their lengths, the image of an interval is an evolution operator
 $I(t)$ acting on the abstract space of states $V$, and functoriality is just a semigroup condition:
\begin{equation}
I(t_1+t_2)=I(t_1) I(t_2), \; \; \;  I(t) \in End(V)
\end{equation}
that may be solved in an almost standard form:
\begin{equation}
I(t)=\exp( t H)
\end{equation}

It is very important to note that there is no Plank constant in Dirac-Segal approach (it is absolutely clear from the explicit solution - Dirac solution - in one-dimensional case). The "Plank constant" actually appears if we consider a
family of theories  $Th_h$, depending on $h$, with the following properties.
For each $h$ there is an algebra $End(V_h)$ is generated by the set of operators  $O(h)_a$, such that
\begin{equation}
[ O(h)_a , O(h)_b ]= h f(h)_{ab}^{c} O_c^h
\end{equation}
and $f(h)$ has a finite limit when $h$ tends to zero, and relations between generators have a finite limit at $h=0$.
Hamiltonian has the following form
\begin{equation}
H(h) = \frac{H^{cl}(h)}{h}, \; \; H^{cl}(h)=y^a(h)O_a(h),
\end{equation}
and coefficients $y^a(h)$ are regular at $h=0$.
Let us call family of observables $A(h)$ having a classical limit if it can be expanded through the generators with
coefficients $z^a(h)$
\begin{equation}
A(h)=z^a(h) O_a(h)
\end{equation}
and $z^a(h)$ are regular at $h=0$.
Then, at $h=0$ symbols $O_a(0)$ form a commutative associative algebra, its spectrum may be considered as a Poissonian
manifold (may be with singularities), its Poisson structure is determined by
\begin{equation}
\{ O_a(0), O_b(0) \}_f(0)=f_{ab}^{c}(0) O_c(0)
\end{equation}
and evolution is given by classical mechanics equation
\begin{equation}
\frac{dA}{dt}=\{H^{cl}(0),  A(0) \}.
\end{equation}

{\bf Warning.} Note that there is no $h$ in the basic evolution equation, the denominator $h$ appears only when evolution operator
is written in terms of operators that have regular classical limit, i.e.
\begin{equation}
\exp(tH)=\exp (\frac{H^{cl}}{h}).
\end{equation}

{\bf Example: Fuzzy sphere}.

Consider $N+1$ dimensional representation of $su(2)$,
\begin{equation}
\{T_a, T_b\}=\epsilon_{abc} T_c,  \;  \; \sum_{a=1}^3 T_a^2=N(N+1).
\end{equation}
Consider generating set
\begin{equation}
O_a=T_a/N
\end{equation}
Then generators $O_a$ form an algebra
\begin{equation}
\{O_a, O_b\}=\frac{1}{N} \epsilon_{abc} O_c,  \;  \; \sum_{a=1}^3 O_a^2=1+\frac{1}{N}.
\end{equation}

If we consider $h=1/N$, then it is a particular case of fuzzy sphere.

\section{Feynman approach and Feynman geometry}
The alternative to Dirac-Segal approach is the Feynman approach.  This approach not just preserves but actively use the basic notions of classical physics such as fields and Lagrangians. The only modification is the principal of minimal (extremal) action. The condition of extremality is replaced by functional integral.

However, such an integral in most cases does not exist as a mathematical object. In early days of functional integral these difficulties were considered as temporary ones, it was believed that heuristic manipulation could produce correct results, and mathematical justification would come later. However, in the last 65 years nobody provided such justification, mathematicians do not even consider it as a problem, and physicists do not want to think about it. We think that the main
problem is that integral has to be taken over infinite-dimensional space, and such integrals rarely exist.

The problem can be easily pointed out in perturbative approach. Infinite dimensional space of integration in most cases
leads to divergences.

The standard way out is to make regularization and then renormalization. Regularization basically replaces the space of functions on the classical space by another space, like space of function on the set of points (vertexes of auxillary lattice) or by space of functions with limited derivative (momentum cut-off). In the standard approach such replacement is often considered as a technical tool, a trick for computation of ideal infinite-dimensional integral (that does not exist). In dreaming about such ideal integral it is mostly assumed that the property of the regularized integrals should be similar to properties of ideal integrals, or at least the difference in properties would somehow go away when regularized space of function would tend to original infinite dimensional one.

In this way of thinking the ideal action often has geometrical meaning formulated in terms of conventional differential geometry, and has symmetries prescribed by differential geometry. People never even ask about what kind of geometry corresponds to the regularized space of function.

Actually, the regularized space of function does not even form an associative algebra. Consider as an example the space of function on a circle of length $2\pi$. Let us make a momentum cut-off, restrict functions to be liner combinations
of $\exp(kx)$ with $|k|< 3N$. Consider the following three functions:
\begin{equation}
f=\exp(2Nx), \; \;  g=\exp(2Nx),     \; \; h=\exp(-2Nx)
\end{equation}
and let us simply restrict the multiplication on the reduced space.
Then multiplying them in one way we get
\begin{equation}
(f \cdot g) \cdot   h=0 \cdot h =0
\end{equation}
since the product of $f$ and $g$ before momentum cut-off equals to $\exp(4Nx)$, is outside the allowed momentum
range and has to be considered as zero. At the same time multiplying in other order
\begin{equation}
f \cdot  (g \cdot  h)=f \cdot 1 = f
\end{equation}
so we have a violation of associativity. It was actually expected since the space of functions of momentum higher than $3N$ does not form an ideal (actually, it does not even form an algebra).
Therefore, according to traditional setup the regularized space of function does not form an algebra, and, therefore, there is no geometrical object that may be attached to it through conventional algebra-geometric correspondence. In particular, all properties and symmetries of the classical action are violated and we do not have any control of this violation, we may only hope that they will be recovered after cut-off of the renormalized theory would be taken to infinity.
We think that such complete loss of geometry in regularized theory means loosing control of symmetries and cannot be a part of proper definition of functional integral.

Situation of lattice regularization is more tricky but the basic phenomena persist. The algebra of functions on vertexes is an honest commutative  and associative algebra, but this is not enough to construct the action. What is actually enough
(as we will show in the following section) is the de Rham DGA (differential graded algebra). It is a supercommutative associative algebra. In the regularization this algebra should be replaced by an algebra of cochains. However, there is no supercommutative associative algebra structure on the space of cochains.

It is the moment of truth. In order to save the idea of functional integral we have to find some kind of algebraic structure on
the regularized space of differentials forms (it does not matter, with momentum or lattice cut-off). This structure should
go to supercommutative DGA structure in the limit when we go to ideal "continuum" limit. One may expect that algebro-geometric correspondence would somehow be extended to this structure, and we call geometry, corresponding to
such structure Feynman geometry.

The good news is that such algebraic structure already exists in mathematics, and it is called $A_{\infty}$ structure.
The brief outline of this structure and its relation to BV (Batalin-Vilkovisky) language is presented in Appendix.

Let us call the  $A_{\infty}$ structure with operations belonging to the trace class {\bf Feynman geometry}.
Here we may consider two types of Feynman geometry. First type is {\bf strong} Feynman geometry when vector space of
$A_{\infty}$ structure is finite dimensional. The second type is  {\bf weak} Feynman geometry, when vector space is
infinite-dimensional but operations are of trace-class, so there is no divergences in Feynman diagrams computations.

Thus, our proposal is to develop Feynman approach to QFT by considering quantum field theories over
 Feynman geometry instead of continuum geometry.

The deviations of Feynman geometry from conventional geometry (as we will see below) happen in two directions. Feynman geometry can be nonsupercommutative but associative DGA, or Feynman geometry may include higher operations. General Feynman geometry deviates from conventional in both directions.

Therefore, our proposal is not to concentrate on each particular example of Feynman geometry but rather consider
QFT over general Feynman geometry. Then, consider the variation of QFT when Feynman geometry is changing - a kind of connection in the bundle of theories over parameters of Feynman geometry, and, finally, define a Feynman QFT over continuum geometry as a limit of QFT over Feynman geometry tending to continuum limit (kind of universal renormalization).

Before we come to examples of Feynman geometry we would like to stress that our world is actually described by Feynman geometry. Experiments can only measure that our geometry is continuum to some accuracy, and they can only improve this accuracy. If QFT may be defined over Feynman geometry we may only ask why in our world this geometry is so close to continuum, but this question looks like the question why cosmological constant is so small, and cannot be answered with the present knowledge of physics.

\section{All actions of traditional QFT written in terms of de Rham DGA}
\subsection{Pure gravity}
The fundamental description of gravity is given by the following data - D-dimensional smooth manifold with a principal $Spin(D)$ bundle
$S$, equipped with the connection $\nabla$ and a morphism $e$ from the tangent bundle to the vector bundle $L$ (Lorentzian bundle,
that is a vector bundle associated to $S$ by D-dimensional (vector) representation of $Spin(D)$).
In this way $e$ may be considered as a $L$-valued one-form, and connection $\nabla$ also may be rewritten in terms of the 1-form $\omega$ with values in the second external power of $L$ (in the case of trivialized bundle $L$)
\begin{equation}
\nabla=d+\omega
\end{equation}
Thus, the Einstein-Cartan action
\begin{equation}
S^{EC}=\int ( e^{a_{1}}  \ldots e^{a_{D-2}} (d \omega^{a_{D-1}a_{D}} + \omega^{a_{D-1}b} \cdot \omega^{b a_{D}}) \epsilon_{a_{1}\ldots a_{D}}
\end{equation}
may be written in the language of de Rham DGA.
\subsection{Chiral fields in 2d and self-dual Yang Mills}
By chiral fields we mean $b-c$ or $\beta-\gamma$ systems. The easiest thing to modify is the spin 0 chiral system  with the action
\begin{equation}
S^{\chi}=\int P \bar{\partial} X=\int P dX
\end{equation}
where $P$ is a $(1,0)$ form.
We would like to express subdivision of differential forms into (p,q)-types using only
$e$-field. To do this we propose to consider $P$ field as a composite field and rewrite it as
\begin{equation}
P=e^{+} p
\end{equation}
where
$e^+$ is and eigenvalue $+i$ eigenvector of $e^a$, $a=1,2$ with respect to $SO(2)$ rotation, and $p$ is a complex scalar field. Thus, the action written in terms of DGA (with an integral) looks as follows
\begin{equation}
S_{DGA}^{\chi}=\int p e dX
\end{equation}
The action of self-dual Yang-Mills may be rewritten in a similar fashion. Namely,
standard action of self-dual Yang-Mills has the form
\begin{equation}
S_{SDYM}=\int Tr P (dA+A^2)
\end{equation}
where $P$ is a self-dual 2-form with values in adjoint representation of the gauge group, $A$ is the connection form in the gauge bundle (we assume that the gauge bundle is trivial and therefore connection may be considered as a 1-form with values in adjoint representation of the gauge algebra.
Like in the chiral field case we will consider field $P$ as a composite field, namely,
we introduce the field $p_{ab}$ with values in adjoint representation of the gauge group times the self-dual 2-tensors of $L$, thus
\begin{equation}
P=e^a e^b p_{ab}
\end{equation}
so the action takes the form that is written in the DGA language
\begin{equation}
S_{DGA}^{SDYM}=\int e^a e^b p_{ab} (dA+A^2)
\end{equation}

{\bf Remark 1}.
If we will not be interested in dynamical gravity, i.e.  only consider it as a
fixed background, we can use just a two-form $E^{ab}$ with values in the antisymmetric self-dual tensors of $L$.
Then, the action would take the form
\begin{equation}
S_{DGA,E}^{SDYM}=\int E^{ab} p_{ab} (dA+A^2)
\end{equation}
It is interesting to note that the $E$ field background is more general than $e$ field since not all $E$ are decomposable into a product of two $e$. It opens a previously unexplored possibility to study self-dual Yang-Mills and (as we will see later) ordinary Yang-Mills theory in a novel background.

{\bf Remark 2}.
An alternative approach would be to introduce the new structure in DGA, a polarization of middle degree elements, that is decomposition of this subspace into a sum of two spaces, invariant under multiplication over degree zero elements. Thus, we may consider the first order actions as actions already written in the language of polarised DGA. Thus, actions (3) and (6) may be considered as written already in terms of polarised DGA. The disadvantage of this approach is that it is not clear how to write Einstein action on the space of polarisations.

\subsection{Spinor fields in arbitrary dimensions}
The very notion of spin connection comes from the desire to couple spinors to gravity.
Therefore it is not surprising that there is such a coupling. What we want to stress, is that this coupling also is written in terms of DGA. Namely, spinor field should be considered as a section of the spinor bundle associated with the $Spin(D)$ bundle,
that basically means that for the case of trivial bundles spinor fields $\psi_{\alpha}$ are zero forms with values in the spinor representation of $Spin(D)$.
The action for the Dirac spinor reads
\begin{equation}
S=\int  \bar{\psi}_{\alpha} ( d  \psi_{\beta} +\omega_{\beta}^{\delta}\psi_{\delta}) \gamma_{a_1}^{\alpha \beta} e^{a_2} \ldots e^{a_D} \epsilon_{a_1 \ldots a_D}+m \bar{\psi}_{\alpha} \psi_{\alpha} vol
\end{equation}
where
\begin{equation}
vol=  e^{a_1} \ldots e^{a_D} \epsilon_{a_1 \ldots a_D}
\end{equation}
and $m$ is the mass of the fermion field.
Note the similarity of this action to action for two-dimensional chiral field - really,
this is how chiral fermions are described in $D=2$.
Similarly one can write the coupling of the spinor field to the gauge field.

\subsection{Sigma-models}
There are two ways to write a free scalar field action in the language of
DGA.

First way works only in dimension 2 and utilises the previously written action for a chiral field. We consider two copies of chiral action (actually, chiral and anti-chiral)
and add a standard $P\bar{P}$ term that now looks as  $p\bar{p}e^+ e^-$ term.
This procedure may be generalized to the general Riemannian metric and a $B$-field
background using approach of \cite{Losev-Marshakov-Zeitlin}.

The second way works in any geometry of the world-sheet and target, and is based
on the first order representation of the D-dimensional sigma-model.
Consider the field $P_{\mu}$ that is a $(D-1)$ -form with values in the pullback of the cotangent bundle to the target manifold. Consider the action
\begin{equation}
S=\int (P_{\mu}dX^{\mu} + P_{\mu} * P_{\nu} G^{\mu \nu}(X))
\end{equation}
Integrating the field $P$ out we get the standard sigma-model action. The first term
looks perfect from the DGA language point of view, however, the second does not -
it involves the Hodge star operation that cannot be formulated in the DGA language.
To deal with this problem we have to go to composite fields like we did in the case of chiral field, namely, we use a field $p_{a_1 \ldots a_{D-1}}$ that is a section of the
$D-1$ external power of $L*$, and make a $D-1$ form out of it using $e$-fields
(taking a pullback to under the maps between $L^*$ and $T^*$):
\begin{equation}
P_{\mu}=e^{a_1} \ldots e^{a_{D-1}}p_{a_1 \ldots a_{D-1}, \mu}
\end{equation}
Then action (12) becomes a legal expression in DGA language, i.e.
\begin{equation}
S=\int e^{a_1} \ldots e^{a_{D-1}}p_{a_1 \ldots a_{D-1}, \mu} dX^{\mu}+
p_{a_1 \ldots a_{D-1}, \mu}p_{a_1 \ldots a_{D-1}, \nu} G^{\mu \nu}(X)) vol(e)
\end{equation}
where $vol(e)$ was defined in equation (11).
It is clear that two ways of writing down sigma-model in the first order formalism are classically equivalent, but there is question if they are really equivalent on the quantum level.
\subsection{Yang-Mills theory}
The Yang-Mills theory in arbitrary dimension can be written along the second way.
Namely, we write down the first order action
\begin{equation}
S=Tr (P(dA+A^2)) + g_{0}^{2} Tr (P*P)
\end{equation}
where $P$ is the $(D-2)$-form with values in the adjoint representation.
Similarly to sigma-model case this form can be expressed through the section of the
$(D-2)$-external power of $L*$
 \begin{equation}
P(p,e)=e^{a_1} \ldots e^{a_{D-2}}p_{a_1 \ldots a_{D-2}}
\end{equation}
and Yang-Mills action takes the form
\begin{equation}
S=\int Tr (P(p,e)(dA+A^2))+Tr( p_{a_1 \ldots a_{D-2}}p_{a_1 \ldots a_{D-2}})vol(e)
\end{equation}
The case of $D=4$ is distinguished since there is another representation of the
Yang-Mills similar to the first way of describing the $D=2$ sigma model.
Namely, in this case we just add to the action od the self-dual Yang-Mills the
$E^2 p^2$ term, namely
\begin{equation}
S^{YM}=\int E^{ab}  Tr (p_{ab}(dA+A^2)) +g_{0}^2 Tr( p_{ab}E^{ab}p_{ce}E^{ce}).
\end{equation}

In order to have standard interaction with gravity we need to consider
$$E^{ab}=e^a e^b.$$
\section{Noncommutative Feynman geometry of the DGA type}
It seems impossible to have a family of finite dimensional supercommutative  DGA that have a de Rham DGA as a limit (may be is possible to have a rigorous proof of this statement but we do not know such a proof). Therefore, to have finite dimensionality we should sacrifice either supercommutativity or associativity (associativity would be replaced by homotopical associativity). In this section we explain what could be done if we sacrifice supercommutativity condition,  case of homotopical associativity will be treated in the next section.

We know two apparently different way to get nonsupercommutative finite dimensional DGA that are close in some sense to de Rham DGA. The first example is quite known DGA that people use to explain the ring structure on cohomology. We call it lattice DGA. The second is related to noncommutative geometry of the manifold.

\subsection{Lattice type}
Consider simplicial complex. Physicists often consider it as a triangulation of a manifold. Differential forms on the manifold may be considered as cochains of simplicial complex since they can be integrated against simplexes.

 Consider the set of simplicial complexes obtained by subsequent barycentric subdivisions. Given two smooth differential forms we can always find such a subdivision that would produce different cochains. Therefore, we  would like to equip cochains on simplicial complex with the structure of differential graded algebra. The question of taking a limit when the number of barycentric subdivision is going to infinity and its relation to de Rham DGA is subtle and we will discuss it elsewhere.

It is easier to discuss the chains rather than cochains, and to give a formula for comultiplication. The problem of finding of supercommutative coassociative comultiplication is known as a Kolmogorov problem and is known to have a negative answer.

 It is enough to study the case of a complex associated to a single  ordered n-simplex.
The complex is a set of linear combinations of basic elements that are  increasing sequences of integers between 1 and n.  We will
denote them as $(n_1, \ldots, n_k)$. The boundary operator $\partial$ acts as
\begin{equation}
\partial (n_1, \ldots, n_k)= - (n_2, \ldots , n_k)+\sum_{j=2}^{k} (-1)^{l} (n_1, \ldots, n_{j-1} n_{j+1} \ldots n_k)+(-1)^{k} (n_1, \ldots, n_{k-1})
\end{equation}
The associative comultiplication reads as

\begin{equation}
\partial (n_1, \ldots, n_k) = \sum_{j=1}^{k} (n_1, \ldots, n_j) \otimes (n_j, \ldots, n_k)
\end{equation}

The associativity of this comultiplication can be explicitly checked, but it is definitely not the cocommutative multiplication.

In the simplest example of 1-dimensional triangulation, namely, let us subdivide a circle into set of intervals.
Let us take orientation on the circle and order end points belonging to each interval according to this orientation.
Then we have comultiplication on chains that induces multiplication on cochains. In the continuum limit this multiplication tends to standard de Rham multiplication, however, for the small size of the interval (let as call it $a$), the nonsupercommutativity of multiplications is reflected in the commutator
\begin{equation}
f \cdot \omega -  \omega \cdot f = \omega a \partial_{x} f
\end{equation}

\subsection{Fuzzy supersphere}

Consider polynomials of two even and two odd variables $C[z_0,z_1,\theta_0,\theta_1]$, where $z$ are even and $\theta$ -odd.
Assign to all of these variables grading 1, and consider a subspace $P_N$ of $C[z_0,z_1,\theta_0,\theta_1]$ of polynomials of degree $N$, i.e. define an Euler vector field $E$
\begin{equation}
E =\sum_{i=0}^{1} \theta_i \frac{\partial}{\partial \theta_i}+z_i \frac{\partial}{\partial z_i}
\end{equation}
then

\begin{equation}
(E-N)P_N=0
\end{equation}
Consider standard $\sigma$-matrixes $\sigma_a$ orthonormal basis in traceless hermitian operators on $C^2$).
Note, that the following operators commute with $E$ and therefore act on $P_N$:
Operators of noncommutative coordinates $x$
\begin{equation}
x_{a}=1/N  \sum_{i,j}  \sigma_{a,ij} z_i \frac{\partial}{\partial z_j}
\end{equation}
Noncommutative de Rham operator
\begin{equation}
D= \sum_{i=0}^{1} \theta_i \frac{\partial}{\partial \theta_i}
\end{equation}
and operators of supercoordinates
\begin{equation}
\psi_{a}=1/N  \sum_{i,j}   \sigma_{a,ij}     \theta_i \frac{\partial}{\partial z_j}
\end{equation}
Note, that operators of supercoordinates are commutators of de Rham operator and operators  of coordinates
\begin{equation}
\psi_{a}=[ D, x_{a}]
\end{equation}
It is clear that the algebra generated by $x$ and $\psi$ is a subalgebra of $End(P_N)$ and thus is finite dimensional.
Moreover, like in case of fuzzy sphere, in the limit $N \rightarrow + \infty$, this algebra tends to the
de Rham algebra on the two-dimensional sphere.

Note,  not only $x$ fail to commute just like in example of the ordinary fuzzy sphere
(with the standard poisson bracket), but also $x$ and $Dx$ are not commuting:
\begin{equation}
[ x_a, Dx_b ]=1/N \epsilon_{abc} Dx_c
\end{equation}
while  supercoordinates supercommute
\begin{equation}
[ Dx_a , Dx_b ]=0
\end{equation}

It would be interesting to find proper generalization of the superfuzzy sphere construction - like fuzzy sphere can be
generalized to finite-dimensional algebra of operators acting on sections of the very ample line bundle on a Kaeler manifold.

\section{Strong Feynman geometry of A-infinity type on a subcomplex}

\subsection{Reminder on induction of A-infinity structure on a subcomplex from DGA on the complex}

In this section we consider the procedure that produces Feynman geometries from infinite dimensional DGA's with finite dimensional cohomology.  This construction was developed in parallel in mathematical physics and in mathematics with different motivation and in different terms, but the construction is basically the same.

Mathematicians were mostly interested in inducing the algebraic structure on cohomology, and they found that besides the ring structure that comes from restriction of multiplication to cohomology (that everybody knows) there are also higher multiplications (Massey operations). If we combine these higher operations with multiplications we get the $A_{\infty}$ structure on cohomology with zero differential. Later mathematicians generalized these constructions, and they start talking about $A_{\infty}$ structures obtained by contraction of the acyclic subcomplex.

In mathematical physics people studied solutions to BV master equations. In particular, they studied solutions to
classical limit of BV equation. As it is well-known, BV action may be considered as a polyvector field. In the special case when BV action is a vector field classical BV equations lead to either $L_{\infty}$ or $A_{\infty}$ structure. From this perspective the procedure of contraction of acyclic subcomplex get the following "physical" interpretation.

The main property of solutions to BV equation is the following. If we decompose BV variables into background variables and
dynamical variables, and integrate out (against a Lagrangian submanifold) dynamical variables we will get the effective action for background variables, and this effective action would also satisfy BV equations.

Therefore, if DGA contains contractible subcomplex as a direct summand (as a subcomplex) we may consider it as
dynamical variables, and treat the rest as a background. One can show that homotopy (that inverts differential on contractable subcomplex) determines the Lagrangian submanifold, and variables of this subcomplex may be integrated out. The effective action would be a sum of Feynman diagramms, and in defining induced $A_{\infty}$ structure we have
to take only tree diagrams. The trivalent vertices in this diagrams correspond to multiplication in the algebra and
propagator is given by homotopy.

Let us reformulate this in the mathematical language.
Suppose that there is a DGA structure on the space $V$ with differential $D: V \rightarrow V$ and multiplication
$ m_2 : V \otimes V \rightarrow V $. Suppose that $V$ has a decomposition as a complex
\begin{equation}
V=V_{IR} \oplus V_{UV},
\end{equation}
with inclusion
 $$\iota : V_{IR} \rightarrow V $$
and projection
 $$\pi : V \rightarrow V_{IR} $$
and homotopy $h$
$$
h: V \rightarrow V
$$
with the properties
\begin{equation}
D h + h D=1 -  \iota   \pi
\end{equation}
(note that the right hand side is the projection on $V_{UV}$)
\begin{equation}
h^2=0
\end{equation}

Then there are induced operations
$$
M_k^{IR} : V_{IR}^{\otimes k} \rightarrow V_{IR}
$$
given as a sum over rooted flat trees with $k$ leaves and one root. For the exact formula reader may consult \cite{Mnev},
but we will give the simplest cases:

\begin{equation}
M_2^{IR}(u,v)= \pi m_2 (\iota u, \iota v)
\end{equation}
\begin{equation}
M_3^{IR}(u,v,w)= \pi  m_2( h   m_2 (\iota u, \iota v), \iota w) + \pi m_2  (\iota u , h m_2(\iota v, \iota w))
\end{equation}

\subsection{ A-infinity algebras of Hodge type as strong Feynman geometry}

Suppose $X$  is a compact manifold with a metric. Then for any positive number $\Lambda$  there is a Feynman geometry constructed as follows. Let us start with de Rham DGA, and decompose it into direct sum of subcomplexes
$ \Omega_E$
corresponding to Laplacian eigenvalues. Let us form $V_{UV}$ from subcomplexes with eigenvalue larger than $\Lambda$ , and $V_{IR}$ from subcomplexes with eigenvalues smaller than $\Lambda$.
\begin{equation}
\Omega^{*}=( \oplus_{E\leq \Lambda} \Omega_E)  \oplus ( \oplus_{E\geq \Lambda}\Omega_E)
\end{equation}

The space
$$ V_{IR}= \oplus_{E\leq \Lambda} \Omega_E$$

 is definitely finite dimensional, and the space
$$  V_{UV}= \oplus_{E\geq \Lambda}\Omega_E$$
is acyclic. In particular, we may take as a homotopy
\begin{equation}
h=d^*/E
\end{equation}

Thus we have Feynman geometry depending on $\Lambda$ and we approach de Rham geometry as $\Lambda$ goes to infinity.

 The parameter $\Lambda$ resembles a momentum cutoff, however, there is a fundamental difference between Feynman geometry approach and old fashioned cutoff approach.
In terms of algebraic structures "old cutoff" means that we are using the algebra with
nonassociative multiplication, and therefore we are breaking many symmetries like gauge symmetry (just recall that to protect gauge symmetry people invented such things as Pauli-Willars regularization). In the Feynman geometry approach
we should properly take into account higher multiplications and write action over a general Feynman geometry.

\subsection{Strict A-infinity algebras of lattice type as strong Feynman geometry}
In section () we studied noncommutative associative algebra of cochains associated to simplicial complex.
However, it is possible to get supercommutative $A_{\infty}$ structure based on decomposition
\begin{equation}
\Omega^*=Ker I \oplus Cochains
\end{equation}
In order to do this we need to find an embedding of cochains into differential forms compatible with the differential.
This problem was solved by Whitney - and corresponding forms are called Whitney forms.
In particular, in the one dimensional case the Whitney forms are
\begin{equation}
C_0 \rightarrow \omega_0(C_0)=C_0(A)(1-t)+C_0(B)t
\end{equation}
\begin{equation}
C_1 \rightarrow \omega_1(C_1)=C_1(BA) dt
\end{equation}
It is quite instructive to write $Ker I$ explicitly.
In degree 1 it consists of 1 forms $\omega(t)dt$ with zero integral :
\begin{equation}
\int_{0}^{1} \omega(t) dt =0
\end{equation}
and in degree zero it consists of functions
that may be represented as
\begin{equation}
f_{\omega}=\int_{0}^{t} \omega(u) du
\end{equation}
It is clear that $f_{\omega}$ takes zero values at $0$ and at $1$ , thus belongs to Ker I, and actually spans it.
The homotopy operator is just
\begin{equation}
h: \omega \rightarrow f_{\omega}
\end{equation}
 and zero otherwise.

These structures were studied in \cite{Mnev}.

In higher dimensions there is no distinguished homotopy, therefore there are many $A_{\infty}$ structures.

\section{Weak Feynman geometry}
Weak Feynman geometry is an $A_{\infty}$ structure that may be not finite dimensional as a vector space but
whose operations are of trace class. Namely, the $m_k$ should be such that
$$
m_k(v_1, \ldots, v_{k-1}, \cdot)
$$
considered for a given  $v_1, \ldots, v_{k-1}$  as a linear operator from $V$ to $V$ should be of the trace class.
Note, that the standard multiplication in the de Rham complex is not weakly Feynman, say, multiplication by 1
is an identity operator in the Hilbert space that is not of trace class.
\subsection{Weak Feynman geometry of Costello type}
The idea of Costello Feynman geometry is to do the two step procedure. First, modify multiplication to make it trace class. Then, most probably, the modified multiplication would not be associative. But it could, and we will see actually would be homotopically associative, thus providing  $A_{\infty}$ structure.

Let us consider definition of index of an operator as a prototype. Index was actually a supertrace of multiplication by 1,
that did not exist. So, the result of multiplication was smoothed by $$ \exp(-\beta \Delta). $$

It gives the idea to replace $m_2$ of de Rham DGA by
\begin{equation}
m_2^{\beta} (v,u)=\exp(-\beta \Delta)m_2(v,u)
\end{equation}

Then one can show that multiplication would not be associative. However, it could be completed to $A_{\infty}$ structure by introducing the following higher operation.
\begin{equation}
m_3^{\beta}(u,v,w)= \exp(-\beta \Delta)  m_2( \int_{0}^{\beta}  \exp(-t \Delta)   h dt    m_2 ( u,  v),  w) + \pi m_2  ( u ,  \int_{0}^{\beta}  \exp(-t \Delta)   h dt  m_2( v,  w))
\end{equation}

This $A_{\infty}$ structure was discovered by Costello \cite{Costello} and for topological strings by Costello-Li \cite{Costello-Li}.

\subsection{String theory in form of Zwiebach as weak Feynman geometry}

The second example of the  $A_{\infty}$ structure is given by string theory in the works of Zwiebach \cite{Zwiebach}.
Thus, string theory development is not orthogonal to the programm we are proposing but rather can be included into it as
an important example.

\section{Instead of conclusion: open questions}
\subsection{Promotion of actions from de Rham DGA to $A_{\infty}$ structures }

We have shown in section 4 that all actions of classical field theory may be formulated in terms of de Rham DGA.
However, it is not clear a priori how to promote them to general $A_{\infty}$ structure, specifically taking into account
possible noncommutativity of operations and existence of higher operations.
Proper promotion should take into account symmetries of the theory,
thus, rigorously the problem looks as promotion of BV action that includes gauge symmetries to general $A_{\infty}$ structure. There is a well-known case where such promotion always exist, this is the case of so-called $BF$ theories.

Originally they were formulated as non-Abelian gauge D-dimensional theories coupled to a field $B$ that is a $D-2$ form with values in adjoint representation of the gauge group with the action
\begin{equation}
S=\int Tr (B (dA+A^2))
\end{equation}

In terms of $A_{\infty}$ structures this may be reformulated as follows. Consider two $A_{\infty}$ structures. First one would be de Rham DGA (considered as $A_{\infty}$ structure), second - the matrix algebra considered as $A_{\infty}$ structure with zero differential. Take a tensor product of these two structures and form a new $A_{\infty}$ structure.
Then write a canonical action for this structure (see Appendix). The outcome would be exactly the BF theory. The only subtle moment here is the issue of integral and $Tr$ in the action.
Actually, in canonical action for $A_{\infty}$ structure we have pairing with a dual field. In case of de Rham DGA fields dual to differential forms may be represented also as differential forms with canonical pairing replaced by an integral. In the case of matrix algebra the dual object can be also considered as a matrix with pairing given by a trace of multiplication.

This explains how to promote $BF$ theory over general Feynman geometry - we should just replace de Rham DGA by
a general Feynman geometry and write down canonical action.

Therefore, we can cover in this way 2D Yang-Mills with zero coupling constant and 3D gravity.

The most interesting question is whether it is possible to promote other actions from the section 3 to general Feynman geometry.
 If it is not possible in general what is the obstruction? Could it be a new selection rule on classical actions?

If such obstruction may vanish on certain Feynman geometry what is a subclass of Feynman geometry where such obstruction vanish?

These are quite new types of questions in QFT,  we are going to address these questions in the following publications.

\subsection{Dirac-Segal approach to QFT versus Feynman geometry approach and the dream of Ultimate QFT}

In this paper we studied two approaches to QFT, Dirac-Segal and Feynman approaches. One may think that they are
orthogonal to each other. In Dirac-Segal approach the geometry of the space-time is a standard continuum geometry, but there is no Plank constant, and no classical fields and Lagrangians. In the Feynman approach there are classical fields,
Lagrangians and Plank constant, but geometry of the space-time is not classical anymore. It may look that these approaches exclude each other.

However, it is not the case. From our experience in dimensions 1 and 2 we know that there are theories that have description in both approaches, and resulting theory is the same. The best known example is the WZWN
that can be described and actually solved in Dirac-Segal approach through integrable representations of affine lie algebras. At the same time this theory has a functional integral representation.

We may conjecture that the Ultimate Definition of QFT would combine these two paths in the following way - both
geometry and physics would be nonclassical. Namely, we expect that it would be possible to generalize Dirac-Segal axioms to an arbitrary Feynman geometry and it would be the goal of axiomatic development of QFT. At the same time
Feynman integrals over fields on Feynman geometry would serve as a constructive examples of solutions to these new axioms.

\section{Appendix: BV master equation and $A_{\infty}$ structures}
For physicist the proper way to understand $A_{\infty}$ structures is through the BV language.
Consider the superspace (called BV space) of the form $\Pi T^* X $ with coordinates $x^a$  on the base $X$ and $x_a^*$ on the cotangent fiber with opposite parity.
Functions on BV superspace may be considered as
polyvector fields on the  space $X$.
The main operator is the BV operator on functions on BV space
\begin{equation}
\Delta_{BV}=\sum_{a}  \frac{\partial^2}{\partial x^a \partial x_a^*}
\end{equation}
The geometrical meaning of $\Delta_{BV}$ is the divergence of the polyvector field; actually, if we apply it to a vector field
 \begin{equation}
v=x_a^* V^a(x)
\end{equation}
we would get
\begin{equation}
\Delta_{BV} v=\sum_{a}\partial_a V^a
\end{equation}
The main object in BV language is a BV action $S(h)$ - even  function on the BV space with values in formal power series in $h$.
The BV master equation reads
\begin{equation}
\Delta_{BV} \exp S(h)/h =0
\end{equation}
If the BV action represents a vector field , then corresponding vector field is odd and
the classical limit of BV equation states that it is a homological vector field:
 \begin{equation}
Q= V^a(x) \frac{\partial}{\partial x^a} \; \;  Q^2=0
\end{equation}
expanding homological vector field in coordinates we get a set of operations
 \begin{equation}
 V^a(x)=\sum_{k=1}^{\infty}\mu_{a_1 \ldots a_k}^{a} x^{a_1} \ldots x^{a_k}/ k!
\end{equation}
where $ \mu_{a_1 \ldots a_k}^{a}$ is considered as an operation
 $$ \mu_{k} : V^{\otimes k} \rightarrow V$$
written in a particular basis.
Set of quadratic equations on operations $\mu_k$ is called set of conditions for $L_{\infty}$ structure.
The first condition states that  $\mu_1$ is a differential, second - that $\mu_1$ together with
 $\mu_2$ form a Leibnitz rule,
third - that Jacobi equation for $\mu_2$ is true up to a commutator of $\mu_1$ and $\mu_3$ and so on.

Note, that $L_{\infty}$ structure is a structure on supercommutative operations.

However, there is a possibility to encode nonsupercommutative operations in homological vector field.
Namely, let us consider take  $X$ to be $C^{MN^2}$  with coordinates $x_{ \alpha}^{i, \beta}$, that we will
denote simply as $\hat{x}^i$.

Consider the BV action that is still a vector field, but of a very special form
\begin{equation}
A=\sum_{k=1}^{\infty}  m_{i_1 \ldots  i_k}^{j}
 Tr  ( \hat{x^*}_j   \hat{x}^{i_1} \ldots \hat{x}^{i_k})/ k!
\end{equation}

If $N$ is big enough then
conditions that $A$ solves classical master equation is exactly the set of quadratic equations for operations $m$ that determine
$A_{\infty}$ structure.

They start with $m_1$ being a differential, Leibnitz rule for $m_1$ and $m_2$, associativity of $m_2$ up to a commutator of $m_1$ and $m_3$ and so on.

\end{document}